# Rational Materials Design for In Operando Electropolymerization of Evolvable Organic Electrochemical Transistors

*Jennifer Y. Gerasimov, Arnab Halder, Abdelrazek H. Mousa, Sarbani Ghosh, Padinhare Cholakkal Harikesh, Tobias Abrahamsson, David Bliman, Jan Strandberg, Matteo Massetti, Igor Zozoulenko, Daniel T. Simon, Magnus Berggren, Roger Olsson, and Simone Fabiano\**

Organic electrochemical transistors formed by in operando electropolymerization of the semiconducting channel are increasingly becoming recognized as a simple and effective implementation of synapses in neuromorphic hardware. However, very few studies have reported the requirements that must be met to ensure that the polymer spreads along the substrate to form a functional conducting channel. The nature of the interface between the substrate and various monomer precursors of conducting polymers through molecular dynamics simulations is investigated, showing that monomer adsorption to the substrate produces an increase in the effective monomer concentration at the surface. By evaluating combinatorial couples of monomers baring various sidechains with differently functionalized substrates, it is shown that the interactions between the substrate and the monomer precursor control the lateral growth of a polymer film along an inert substrate. This effect has implications for fabricating synaptic systems on inexpensive, flexible substrates.

## 1. Introduction

Organic electrochemical transistors (OECT) have emerged as one of the most widely investigated devices for diverse bioelectronics applications such as neural interfaces, sensors, printed logic circuits, and, more recently, neuromorphic computing devices.[1] While OECTs channels are generally patterned between source and drain contacts by photolithography or printing processes before operation, there is a growing interest in 'in operando' electropolymerization of the conducting polymer channel as a means of fabricating evolvable organic electrochemical transistors (EOECTs).[2] This approach is appealing for constructing artificial synapses for neuromorphic systems where the synaptic weight is represented by the channel conductance because it allows for permanent incremental modulation and independent evaluation of the electrical transistor characteristics in operando.

An EOECT device consists of a pair of opposing metal electrodes thermally deposited on a silicon oxide substrate (**Figure 1**). A solution containing a monomer precursor of a conducting polymer and a supporting electrolyte is contained in a polydimethylsiloxane well. Voltage applied asymmetrically between the two metal contacts drives the oxidative polymerization of a conducting polymer channel between the metal contacts, while a grounded Ag/AgCl pellet electrode provides the counter-reaction for monomer oxidation.

In selecting a monomer precursor for electropolymerization, especially when considering requirements of implantable and biologically interfaced neuromorphic circuits, it is beneficial if the precursors exhibit minimal toxicity and that both device fabrication and evaluation can be performed at low voltages in an aqueous environment. This requires that the monomer is water-soluble, produces a polymer that exhibits good ionic conductivity, can be polymerized at potentials within

J. Y. Gerasimov, A. Halder, P. C. Harikesh, T. Abrahamsson,
M. Massetti, I. Zozoulenko, D. T. Simon, M. Berggren, S. Fabiano
Laboratory of Organic Electronics
Department of Science and Technology
Linköping University
SE-60174 Norrköping, Sweden
E-mail: simone.fabiano@liu.se

A. H. Mousa, D. Bliman, R. Olsson
Department of Chemistry and Molecular Biology
University of Gothenburg
SE-412 96 Gothenburg, Sweden

S. Ghosh
Department of Chemical Engineering
Birla Institute of Technology and Science (BITS)
Pilani Campus, Vidya Vihar, Pilani, Rajasthan 333031, India

J. Strandberg
Printed Electronics
RISE
Research Institutes of Sweden
SE-602 21 Norrköping, Sweden

R. Olsson
Chemical Biology and Therapeutics
Department of Experimental Medical Science
Lund University
SE-221 84 Lund, Sweden

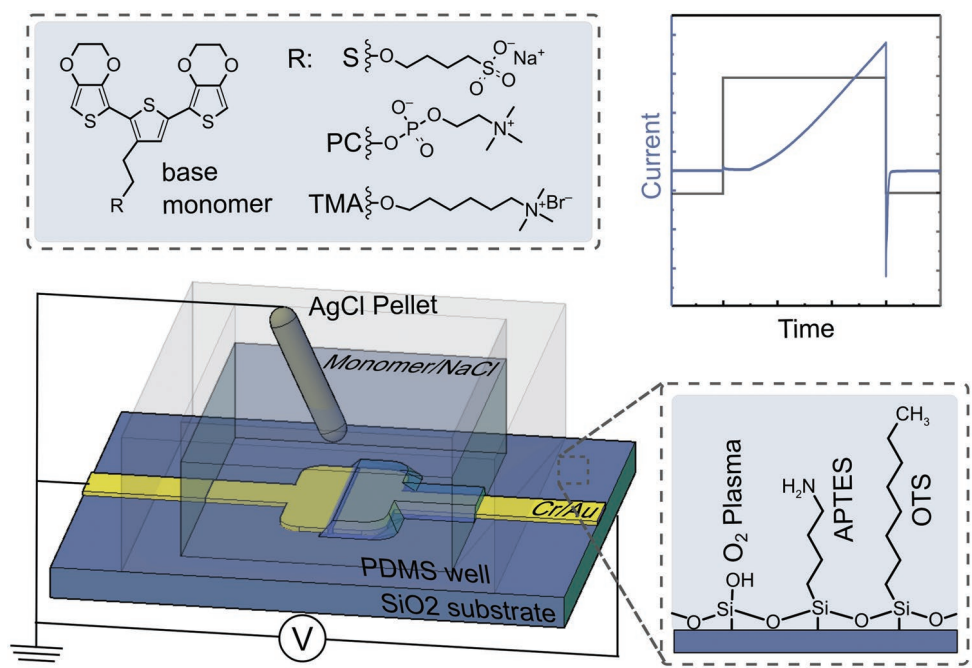

**Figure 1.** Device schematic. Schematic of the device, including the chemistry of the monomer derivatives and the surface modifications used in this study. The monomer is oxidatively electropolymerized by applying a voltage above the polymerization threshold value for a period of time. As the polymer crosses the gap between the source and drain and makes contact with the opposing electrode, there is an increase in the channel conductance, which is evidenced as an increase in the drain current.

the electrochemical window of water, and produces a device that exhibits a peak transconductance at low voltages. Evolvable devices that fulfill these criteria greatly simplify device operation and bypass many of the challenges to integrating this technology with biological systems. The monomer that we introduced for EOECT fabrication in our previous work is composed of a 2,5-bis(2,3-dihydrothieno[3,4-b][1,4]dioxin-5-yl) thiophene (ETE) backbone, which has been functionalized on the central thiophene with a ethoxy-1-butanesulfonic acid sidechain (ETE-S), fulfills all of the aforementioned criteria.[2a,3] ETE-S is soluble in water, an excellent ion conductor is oxidatively polymerized below 0.3 V versus Ag/AgCl, and produces an intermediately doped conducting polymer, making its electropolymerization more biocompatible, energy-efficient, and less complicated than that of the more thoroughly studied ethylene dioxythiophene (EDOT).[2b,c,4]

One difficulty that we encountered in fabricating devices by ETE-S electropolymerization is extreme batch-to-batch irreproducibility in the amount of time required for the polymer to cross the 30 μm gap between the source and drain electrodes. The time required to cross the gap fluctuated from seconds for some batches to never making the connection within 10 min for others. We attributed this effect to nonoptimal interaction of the ETE-S monomer/PETE-S polymer with the surface of the untreated silicon substrate[2a] and sought deliberately functionalize the substrate to better control its surface chemistry and the polymer spreading.

Considering how thoroughly electropolymerization of conducting polymers has been examined,[5] surprisingly, only a few studies have addressed polymer spreading away from the electrode along the inert substrate. The most notable effort was that of Nishizawa and colleagues, who investigated the effect of substrate silanization on the rate of lateral growth and morphology of a polypyrrole film.[6] More recently, a cursory investigation of substrate effects on the morphology of PEDOT and its derivatives was also performed as part of a larger study by Watanabe et al.[7] Given the growing popularity of in situ device fabrication, understanding the spreading of conducting polymer channels on an insulating substrate is essential for the rational design of the next generation organic bioelectronic devices.

Here, we deliberately control the surface properties of the $SiO_2$ substrate by introducing hydrophobic trimethoxy(octadecyl)silane (OTS), partially positive (3-aminopropyl) triethoxysilane (APTES), and partially negative (oxygen plasma treated) functional groups. We postulated that the charged sulfonate sidechain plays a role in determining the interactions with the substrate and sought to test this theory by complimenting the ETE-S, which is predominantly negatively charged at pH 7, with a zwitterionic phosphocholine ETE derivative (ETE-PC) and a positively charged trimethyl ammonium-containing derivative (ETE-TMA). This systematic approach enables us to construct a set of general principles that can be used to rationally adapt EOECTs to novel substrates.

## 2. Results and Discussion

### 2.1. Design and Synthesis of Mixed Conducting Polymers

Maintained aqueous solubility at the concentrations used for the experiments was a prerequisite for any changes made to the monomer. Phosphoryl choline, structurally inspired by the zwitterionic hydrophilic head of phospholipids, was chosen as a substituent to obtain a neutral monomer with high hydrophilicity. A quaternary ammonium substituent was selected to obtain a positively charged monomer. A hexyl linker was chosen to distance the positive charge from the thiophene core of the timer. The hexyl trimethyl-ammonium substituent was previously used to obtain positively charged polythiophenes[8] and polyethylenedioxythiophenes.[9]

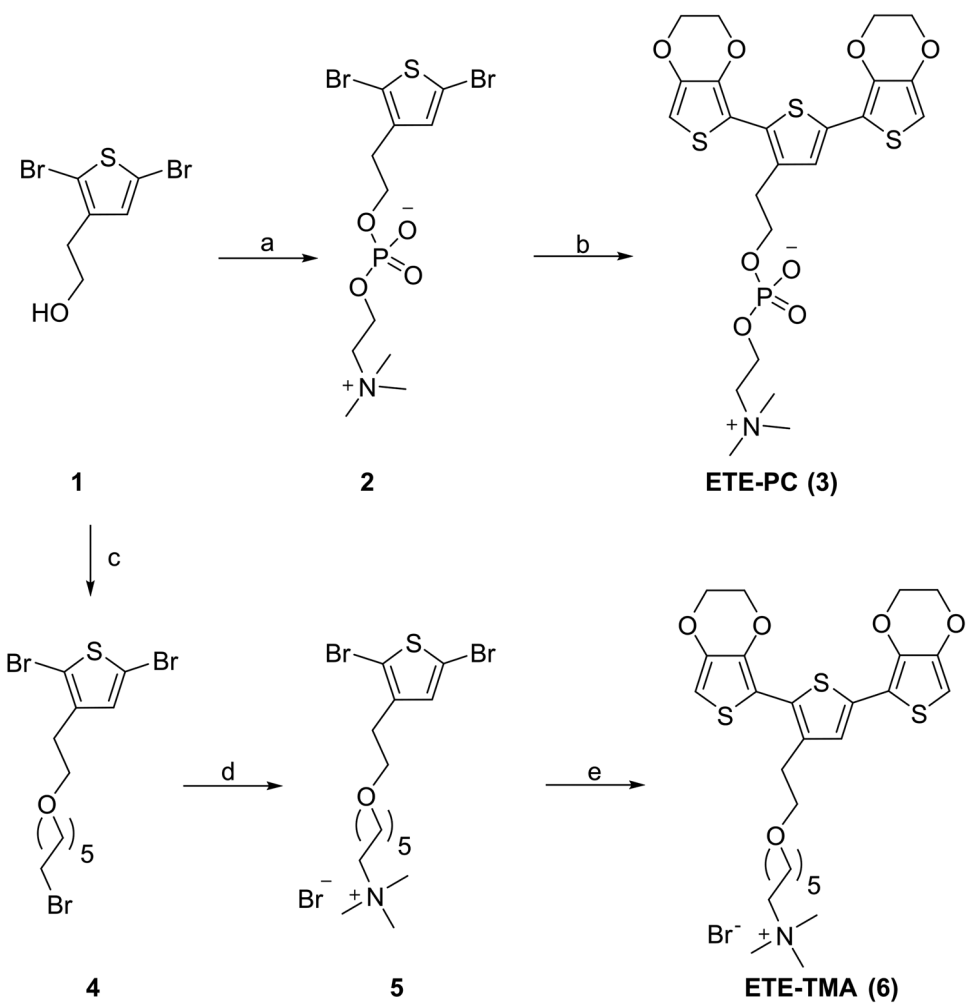

**Scheme 1.** Synthesis route to ETE-PC (3) and ETE-TMA (6). a) i) 2-chloro-1,3,2-dioxaphosphalane-2-oxide (1.02 eq.), Et$_3$N (1.1 eq.) in dry THF, −40 °C to r.t., 7h. ii) Me$_3$N (6 eq.) in CH$_3$CN, 60 °C, 18 h, 57% over 2 steps. b) EDOT-borolane (3 eq.), PEPPSI-IPr (10 mol%), KF (6 eq.) in water:DMF 1:1, 80 °C, 5 h, 66%. c) NaH in dry THF, r.t., 15 min, cooled to 0 °C, 1,6-dibromohexane, r.t., then 65 °C for 16 h, 54%. d) NMe$_3$ in THF, 85 °C, 88%. e) EDOT-borolane (2.2 eq.), PEPPSI-IPr (7 mol%), KF (6 eq.) in water:DMF 1:1, 85 °C, 16 h, 65%.

The synthesis of both phosphoryl choline monomer (ETE-PC, 3) and trimethylammonium monomer (ETE-TMA, 6) starts with 1,4-dibromination of commercially available 2-thiophenyl-2-ethanol (**Scheme 1**) as previously described.[10] Analogous to a procedure for the synthesis of phosphoryl choline substituted ethylene dioxothiophene,[11] the 1,4 dibromo-thiopheenethanol was reacted with 2-chloro-1,3,2-dioxaphosphalane-2-oxide resulting in a phosphalane intermediate which was treated with a trimethylamine solution in acetonitrile at 60 °C, giving the dibromothiophene phosphoryl choline (2) in 57% yield over two steps. To obtain the precursor for the ETE-TMA (6), 1,4-dibromo-thiopheenethanol was alkylated with 1,6-dibromohexane followed by a nucleophilic substitution using trimethylamine as a nucleophile resulting in the dibromothiophenyl ammonium intermediate (5). With access to both dibromo-intermediates (2 and 5), the monomers were obtained using a palladium-catalyzed Suzuki cross-coupling reaction with boronate ester functionalized ethylenedioxythiophene (EDOT-borolane) (Scheme 1). Both ETE-PC and ETE-TMA were characterized by 1H-, 13C-NMR, and high-resolution mass spectrometry (see SI for details) and display absorption bands in UV-vis at 350 nm, characteristic for ETE-monomer systems. The monomers are soluble to at least 10 mg mL$^{-1}$ in aqueous media, well within the range required for channel growth.

### 2.2. Electrochemical and Device Properties of ETE-X Films

The oxidative polymerization of the ETE-X monomers and the electrochemical characteristics of the resulting films were evaluated by cyclic voltammetry (CV) on a fluorine-doped tin oxide electrode (**Figure 2**). The oxidation peak in the first sweep of the cyclic voltammogram in a solution of ETE-X monomer, corresponding to nucleation of the conducting polymer on the FTO electrode, is observed at 0.30 V for ETE-S, 0.47 V for ETE-PC, and 0.54 V for ETE-TMA (Figure 2a). The onset potential of the electropolymerization correlates with the charge on the monomer, with the negatively charged monomer having the lowest onset potential, the positively charged monomer having the highest onset potential, and the zwitterionic monomer having an onset potential with an intermediate value. The underlying cause of this trend must be further investigated, but we can speculate that electrostatic interactions affect either the distance between the monomer and the positively charged electrode or the thermodynamics of the monomer oxidation. The fifth cycle of the CV shows an increase in the capacitive current, extracted from the non-Faradaic current region, by two orders of magnitude upon ETE-X polymerization from ≈40 nA for bare FTO to 4–10 μA for the polymer layer (Figure 2b). The marked increase in the capacitive current for all polymer layers demonstrates that ETE-PC and ETE-TMA can also be electropolymerized and can be described as having a volumetric capacitance, thus making them compatible with EOECT technology.[12]

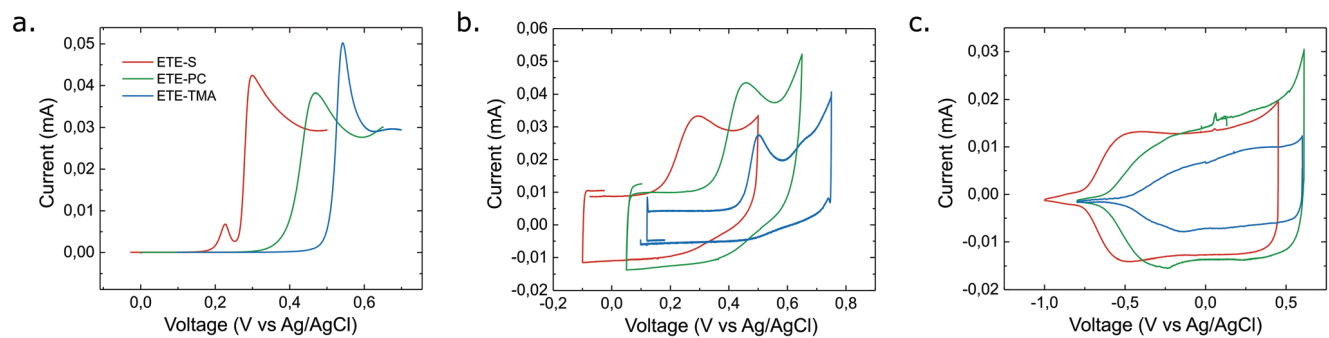

**Figure 2.** Cyclic voltammetry characterization of ETE-X electropolymerization and the resulting PETE-X film. a) The first CV sweep and b) the last CV cycle of a fluorine-doped tin oxide (FTO) electrode in a solution of ETE-X monomer. c) CV characterization of polymer films formed by voltage cycling (5 cycles) evaluated in the absence of ETE-X.

After electropolymerization by CV, the monomer solution was removed and replaced with an aqueous solution containing 100 mM NaCl to isolate the electrochemical properties of the polymer film. The working electrode was cycled until the voltammogram showed no change in subsequent cycles. Cyclic voltammograms over an expanded range were performed in a nitrogen atmosphere to eliminate the effects of the oxygen reduction reaction (Figure 2c). All three ETE derivatives showed CV responses characteristic of p-type conducting polymers, consisting of regions corresponding to the conductive oxidized state and an insulating reduced state.

### 2.3. Molecular Dynamics Simulation

As noted previously, we suspect that the interactions between the monomer and the inert substrate supporting the source and drain electrodes is a determinant of whether lateral spreading away from the active electrode will be observed during polymerization. We thus study the interaction of the monomer precursors with three distinctly functionalized surfaces using molecular dynamics (MD) simulations. As representative systems, we evaluate a partially negative hydrophilic $SiO_2$ surface, as would be obtained by plasma cleaning, alongside a partially positive silanized surface modified with amine-terminated APTES and a silanized hydrophobic surface modified with OTS. A simulation box of ≈115 Å × 115 Å × 100 Å was packed with 200 monomer molecules, the associated counter-ions, and 80% water molecules using the Packmol software package.[13]

The snapshots of the equilibrated MD systems and the corresponding density distributions of the monomers from the surface in the direction out-of-plane from the substrate are shown in **Figure 3**. In all cases, there is a significant degree of aggregation in the solution. The density distribution plots describing the APTES-modified substrate reveal the adhesion of ETE-S and ETE-PC monomers at the substrate surface whereas ETE-TMA does not adhere to the substrate. The extent of monomer adhesion at the APTES surface appears to be driven by electrostatic interactions. The same trend can be observed on the partially negative hydrophilic $SiO_2$ surface. While the anionic ETE-S is predominantly present as aggregates in solution, the ETE-PC and ETE-TMA are mostly distributed at the surface. Visual examination of the final snapshots of monomer interaction with the OTS-modified substrates indicates that all three monomers are incorporated into the OTS layer, which can be expected due to the hydrophobic core of the ETE-S. The density distribution plots show that ETE-S and ETE-TMA penetrate through the hydrophobic OTS layer, which is approximately 26 Å thick in its extended form.[14] While ETE-PC does not appear to penetrate into the OTS layer, it does not exhibit the same gap between the monomer and the substrate as is seen with clearly non-interacting couples like APTES/ETE-TMA. When introduced to an OTS-modified substrate, the ETE-TMA is present in two discrete populations: as aggregates in solution and embedded in the OTS.

### 2.4. OECT Channel Growth

To experimentally study the influence of device substrate chemistry on EOECT channel growth, we modify the silicon oxide substrate prior to depositing the source and drain contacts through either oxygen plasma treatment to obtain a partially negative hydrophilic surface or silanization to obtain a partially positive hydrophilic or hydrophobic surface. The formation of the EOECT channel was evaluated on each of the three modified substrates in combination with each of the three ETE-X derivatives (**Figure 4**).

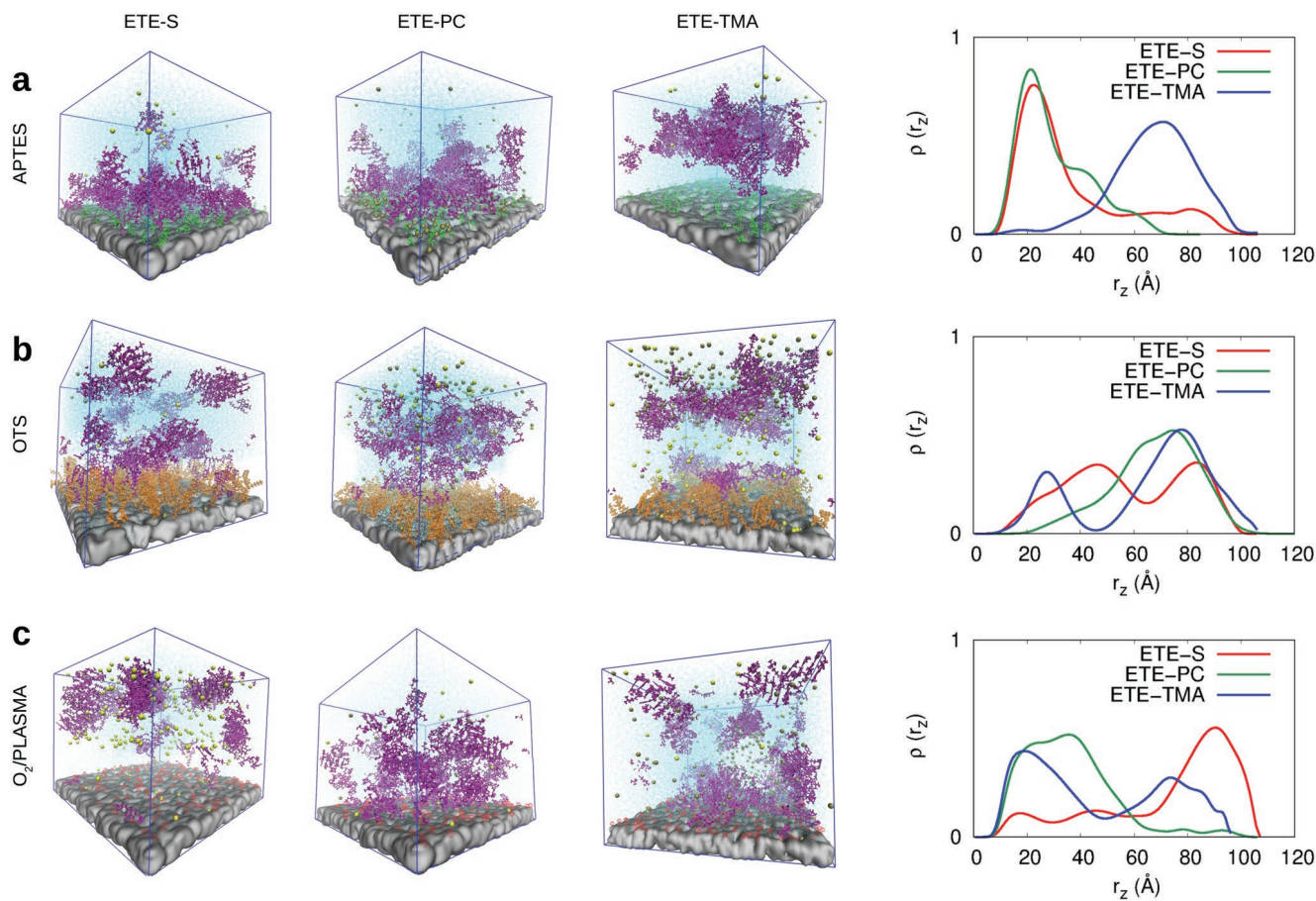

**Figure 3.** Snapshots of the final state of molecular dynamics simulations and the densities of monomers distribution from the substrate in the substrate direction (z) of all the monomers, ETE-S, ETE-PC, and ETE-TMA for all modified $SiO_2$ surfaces, a) APTES, b) OTS, and c) $O_2$/PLASMA.

The ETE-X monomers were electrochemically polymerized at a constant offset potential from the oxidation current peak to generate EOECT channels. This was done to maintain some consistency in the rate of electropolymerization across the monomers. Voltages were selected to be 0.5 V for ETE-S, 0.6 V for ETE-PC, and 0.7 V for ETE-TMA. Channels were grown asymmetrically by applying a constant electropolymerization voltage to one of the source/drain contacts for 60 seconds and measuring the current passing through the channel between the source and the drain (Figure 4a). Optical microscopy images were acquired at intervals of one second and the final images are displayed as an inset in Figure 4a. Time-course analysis of the intensity and spreading are in Figure S24, Supporting Information.

We find that, as suggested by previous results and supported by MD simulation, tailoring the surface properties of a device substrate to the conducting polymer material is essential for the promotion of lateral polymer growth and the formation of the channel between the EOECT's source and drain electrodes. We observe a significant influence of substrate modification on channel formation and device-to-device variability. For ETE-S, facile and reproducible channel formation was observed for both APTES- and OTS-modified substrates, which is evidenced by a steady increase in the current measured between the two gold electrodes. However, for the $O_2$ plasma-treated substrate, neither an increase in the current nor any significant lateral polymer growth is observed within the 60 s timeframe of the experiment. For ETE-PC, a zwitterionic monomer, channel formation was observed on every substrate that was evaluated. For positively charged ETE-TMA, robust channel formation was observed only on negatively charged plasma cleaned substrates, but a small increase in $I_D$ was also observed on APTES-modified substrates.

Results from the MD simulation on cationic and anionic substrates are in perfect agreement with the electrical characterization of channel formation and the microscopic characterization of polymer spreading behavior. Accumulation of ETE-X at the substrate surface leads to the formation of a dense layer with a higher effective monomer concentration in the substrate plane and facilitates the mass transfer and film growth in the lateral direction.

The results from the hydrophobic OTS-modified substrates are somewhat more difficult to interpret in the context of the MD simulations. Whereas the channel formation on hydrophilic surfaces is favorable when a high-density monomer layer is present near the surface, this correlation breaks down on OTS-modified substrates. For example, the monomer distribution profile of ETE-TMA has a well-defined peak near the OTS surface, but ETE-TMA does not form a conductive channel while ETE-PC does form a highly conductive channel but does not exhibit any apparent surface accumulation. One explanation for this behavior is that the ETE-TMA monomers are entrapped in the OTS in quantities that are insufficient to form a continuous layer while the diffusion coefficient of the entrapped monomer is constrained to a value that prevents the ETE-TMA material from forming fibers. The lack of continuity/interaction between the ETE-TMA entrapped in the film and the ETE-TMA in the solution also precludes film formation. Unlike the distribution of ETE-TMA, the distribution of ETE-S and ETE-PC is continuous rather than discrete between the OTS-entrapped species to the solubilized species, indicating that there is potential for interaction between the two.

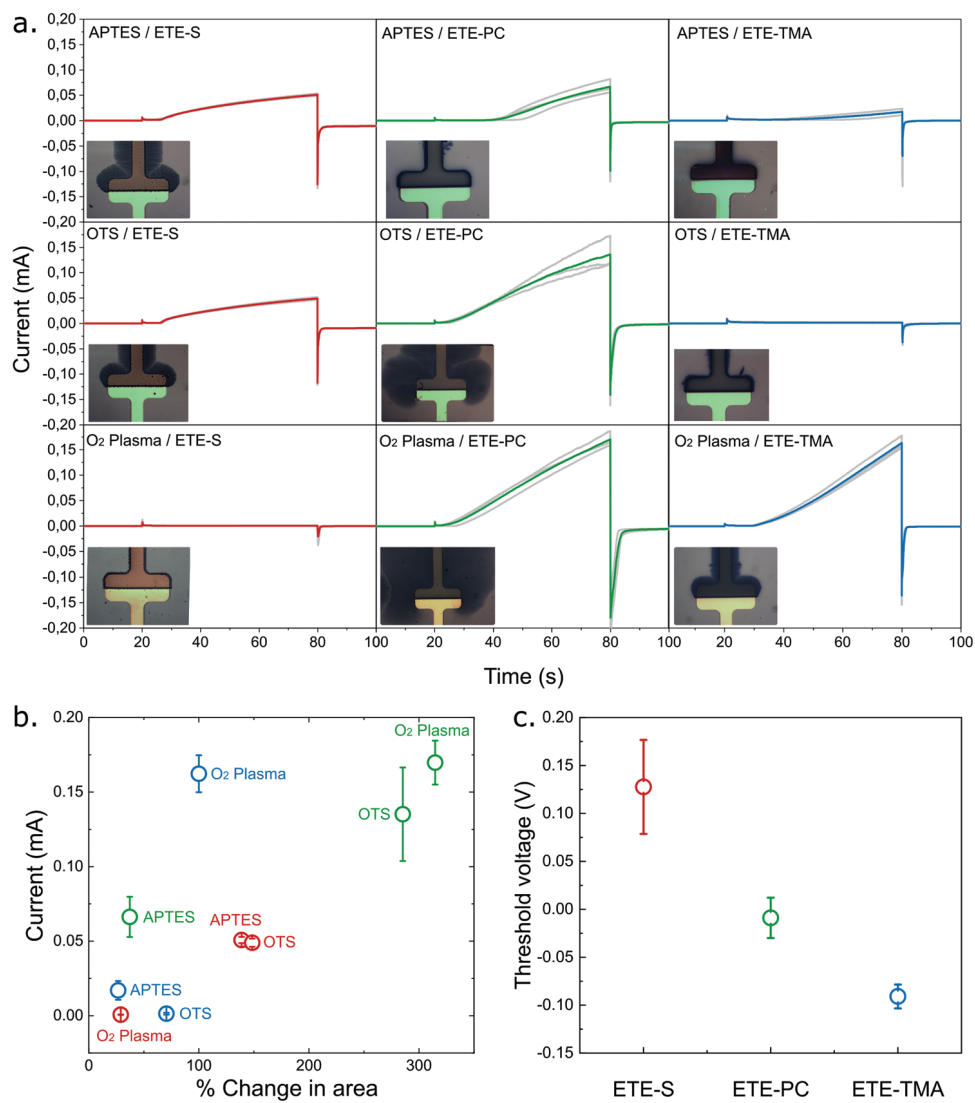

**Figure 4.** Device fabrication, spreading behavior, and electrical characterization. a) The current monitored during EOECT fabrication, with the inset representing the microscopic image at the end of the 60 s deposition step. b) The current plotted against the change in area covered by the polymer, which is a characteristic of polymer spreading, for each combination of monomer and surface modification. c) The threshold voltages of EOECTs fabricated with ETE-S, -PC, and -TMA.

When the final current is plotted as a function of change in area, it is clear that channel conductance correlates with polymer spreading away from the gold electrode (Figure 4b). We postulate that the extent of spreading is determined by electrostatic and hydrophobic interactions between the substrate and the ETE-X monomer. Since the ETE-S monomer consists of a hydrophobic conjugated π-system and a sidechain that is predominantly negatively charged at neutral pH, it has favorable interactions with positively charged APTES-modified substrates and hydrophobic OTS-modified neutral substrates, leading to facile polymer spreading and efficient channel formation on these substrates. On the other hand, channel formation for plasma cleaned negatively charged substrate did not occur within the timeframe of the experiment, presumably due to the electrostatic repulsion between the polymer and substrate. In contrast, ETE-PC spreads best and forms the most highly conductive channels on anionic and hydrophobic substrates. Although ETE-PC has both positively and negatively charged functional groups on the sidechain, interactions between the substrate and the cationic quaternary ammonium group are more favorable due to greater stearic availability when compared to the anionic phosphate group. While PETE-TMA shows less spreading overall than either PETE-S or -PC, the trend of the ETE-TMA series is consistent with electrostatic control of channel formation. There is some discrepancy in that the channel formation is observed on the partially positive APTES. We presume that this current is the result of sedimentation of polymer aggregates in the channel. From the time-course images taken after polymerization is complete, it appears that a fraction of PETE-TMA does not deposit at the surface but rather is present as a dispersion that can then settle in the gap between the source and drain and provide a conductive path (Figure S25, Supporting Information).

The side-chain modification of the PETE-X also affects the transfer characteristics of the resulting devices. Due to poor film adhesion for some monomer-substrate combinations and inconsistency in rinsing protocols, the absolute values of the device characteristics are too variable to make any meaningful comparison (Figure S27, Supporting Information). However, transfer curves show that the threshold voltage becomes more negative with increasing positive charge on the sidechain (Figure 4c). This behavior is beneficial because it allows for the fine-tuning of the transfer characteristics for the specific needs of an application.

### 2.5. Evolvable Transistors on Flexible Substrates

Electronics are currently trending toward non-traditional substrates to accommodate conformable, wearable, and bio-integratable devices.[15] The adaptation of EOECTs to novel substrates must be accompanied by side-chain engineering to match the surface energy between the monomer and the substrate. Here, we adapt the EOECT to screen-printed devices on a polyethylene terephtalate (PET) substrate. The devices are constructed by first depositing a conductive carbon ink to pattern the electrodes and then following with an insulating ink to block the excess electrode surface area. The exposed electrodes forming the EOECT channel nominally have a width of 2000 µm, a separation of 100 µm between the carbon electrodes and a separation of 100 µm between the edge of the transistor channel and the insulating layer. The voltage sequence to electropolymerize the channels was the same as was used for silicon-based devices.

Of the three available monomers, only ETE-PC forms a connection within a minute of electropolymerization time (**Figure** 5a). Taking into consideration both the appearance and transfer characteristics of the printed channels, we can infer that PETE-PC spreads on both the insulator and along with the PET foil, forming a conductive channel between the source and drain (Figure 5b,c). Meanwhile, PETE-S spreads only along the insulator, which produces a marked increase in capacitance (Figure 5, inset) but does not make contact with the opposing electrode. As on silicon-backed devices, the PETE-TMA appears to exhibit poor adhesion to any of the available surfaces, including the carbon electrode, and likely forms a solubilized dispersion. While there is some increase in the capacitance current on the carbon electrode used for ETE-TMA electropolymerization when compared to a bare carbon electrode, this increase is small compared to the PETE-S-modified electrode.

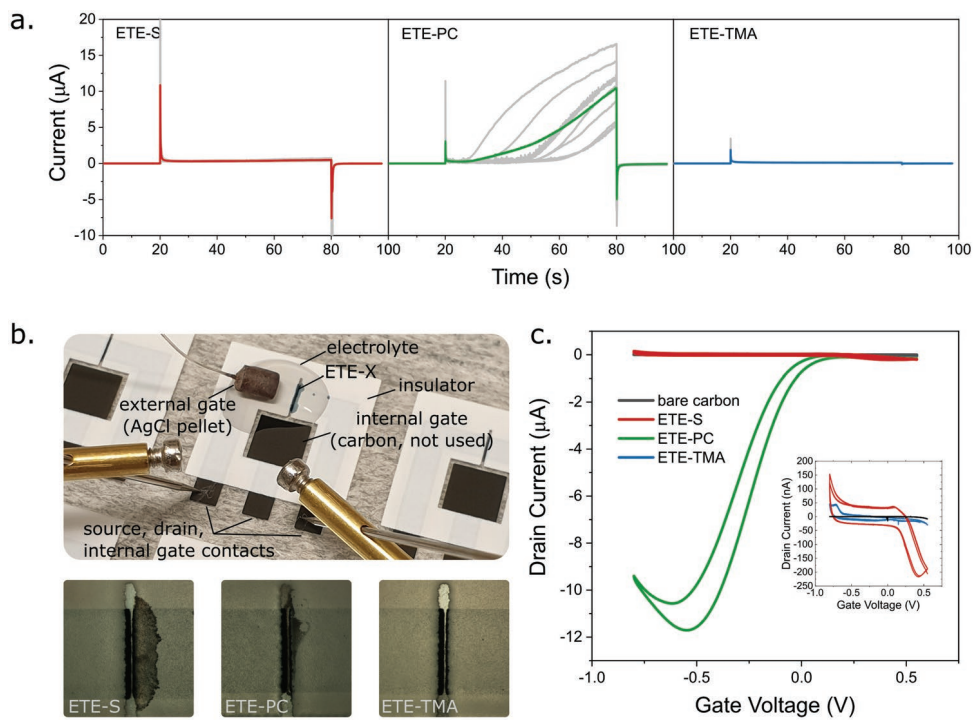

**Figure 5.** EOECT channel formation on screen-printed devices. a) The current monitored during EOECT fabrication. b) The screen-printed EOECT device (top) as well as a microscopic image of the channel (bottom). c) Representative transfer characteristics of devices formed using each ETE-X channel ($V_D = -0.2$) as well as the equivalent voltage sweep of a bare carbon electrode.

## 3. Conclusion

In order for any technology to advance, a thorough understanding of the principles underlying its behavior must be established. Here, we describe a key engineering principle to ensure the success of evolvable OECTs, which have applications in a number of fields, most recently as synaptic devices in neuromorphic computing. We demonstrate, both experimentally and through MD modeling, that the electrostatic and hydrophobic interactions between the substrate and the monomer play a significant role for the lateral growth of polymer film across a substrate. EOECTs based on three differently charged ETE derivatives, negatively charged ETE-S, positively charged ETE-TMA, and zwitterionic ETE-PC were investigated in combination with substrates modified to display different types of surface charge—partially negative oxygen plasma cleaned silicon, partially positive APTES, and hydrophobic OTS. By modifying the surface charge of the substrate and/or the charge on the monomer sidechain, we can rationally tune the spreading of an electropolymerized film as well as the electrical properties of the transistor. This broad understanding of film spreading is important when transferring EOECTs to new substrates, as we demonstrate by growing EOECTs on a flexible PET substrate. More broadly, this understanding will also be essential in applying electropolymerization as a useful tool to modify future micro- and nanobioelectronic interfaces.

## 4. Experimental Section

*Materials Synthesis and Characterization*: ETE-S monomer was synthesized by following previously reported protocols.[16] Details of materials synthesis procedures and characterization for ETE-PC and ETE-TMA monomer could be found in the supporting information.

*Electrochemical Characterization*: Cyclic voltammograms of the ETE-X monomers were obtained in an aqueous solution containing 2 mM monomer dissolved in 10 mM NaCl. A flat FTO-coated glass substrate serves as a working electrode, a Pt mesh electrode as a counter electrode, and an Ag/AgCl pellet electrode as a reference electrode. The CVs of the formed polymer films were obtained in a nitrogen atmosphere.

*Device Fabrication and Surface Modification*: For EOECT device fabrication, silicon substrate with a 1 µm thermally grown oxide layer was cleaned by sequential sonication in 2% Hellmanex, DI water, acetone, and isopropanol and then dried under a stream of nitrogen gas. Clean substrates were then activated by oxygen plasma treatment for 2 min using an input power of 200 Watts. APTES and OTS were purchased from Sigma Aldrich/Merck and stored in nitrogen atmosphere at room temperature. To modify the $SiO_2$ surface with APTES and OTS by vapor-phase deposition, substrates were placed in a stainless steel petri dish and 80 µL of the appropriate silane was dropped at the bottom of the petri dish in such a way that it does not make contact with the substrate. The petri dish was placed on a hot-plate set to 80 °C for 2 h. Excess silane was removed by sonicating the substrates in acetone followed by isopropanol for 5 min each.

After substrate preparation, a set of source and drain electrodes (2 nm Cr, 18 nm Au) with channel dimensions (width × length) of 1000 × 30 µm were thermally evaporated on the APTES-modified $Si/SiO_2$ substrate using an evaporation mask (Source–Drain Deposition Mask for Low-Density OFETs, Osilla Ltd, UK). Electropolymerization of conducting monomers was done using a SP-300 Bio-Logic potentiostat/galvanostat (Bio-Logic Science Instruments, France). Substrates that were not subjected to any treatment besides plasma cleaning were cleaned in a UV-Ozone cleaner for 10 min immediately prior to use. During deposition, a working electrode was connected to the source terminal while the drain was grounded together with an Ag/AgCl pellet electrode (Warner Instruments, USA), which serves both as the gate and a counter-reaction for the electropolymerization. Solutions containing 2 mM of the ETE-X monomer were prepared in DI water containing 10 mM NaCl electrolyte. The voltage used to electropolymerize each monomer is selected to be 0.2 V above the monomer oxidation peak (0.5, 0.6, and 0.7 V for ETE-S, -PC, and -TMA, respectively) and is applied for 60 s. A voltage of −0.2 V is applied before and after the electropolymerization voltage.

Device characteristics like output and transfer curve were obtained on a Keithley Model 4200 Semiconductor Characterization System (Keithley Instruments, USA). An in-plane electrode modified with silver paste was used as the gate.

*Printed EOECTs*: The PET substrate Polifoil Bias is purchased from Policrom Screen. Carbon ink 7102 printing paste from DuPont is used for the electrode contacts. Insulating ink (5018, DuPont) was used for electrode isolation.

Devices comprising a set of carbon source and drain electrodes protected by an insulating layer were screen printed on a PET substrate using methods described previously.[17] Nominal channel dimensions were 2000 × 100 µm ($W \times L$). The conducting channels on printed EOECTs were formed and characterized using a Keithley 2614B System SourceMeter SMU (Keithley Instruments, USA) operated with a custom LabView program (NI, USA).

*Molecular Dynamics Simulations*: The interaction of the three monomers, ETE-S, ETE-PC, and ETE-TMA with the three different substrates were studied using MD simulations. All the MD simulations were run in LAMMPS simulation package.[18] 200 monomer molecules, 80% water molecules, and the required number of $Na^+$ and $Cl^-$ counterions were packed in a simulation box of ≈115 Å × 115 Å × 100 Å using the Packmol package.[13] Bonded and nonbonded interactions were defined using the general AMBER force field (GAFF)[19] as implemented in Moltemplate.[20] ≈10 Å thick amorphous $SiO_2$ substrate was formed using Visual Molecular Dynamics software[21] with the inorganic builder plugin and three types of surface modifications were carried out namely, $SiO_2$-APTES, $SiO_2$-OTS, and $SiO_2$-$O_2$/plasma.

## Supporting Information

Supporting Information is available from the Wiley Online Library or from the author.

## Acknowledgements

J.Y.G. and A.H. contributed equally to this work. This project was financially supported by the Swedish Foundation for Strategic Research (RMX18-0083), the Swedish Research Council (2018-06197), the European Research Council (834677 "e-NeuroPharma" ERC-2018-ADG), and the Swedish Government Strategic Research Area in Materials Science on Functional Materials at Linköping University (Faculty Grant SFO-Mat-LiU 2009-00971). The authors also acknowledge financial support from the Knut and Alice Wallenberg Foundation and the Önnesjö Foundation. Part of the study was accomplished within MultiPark, and NanoLund —Strategic Research Areas at Lund University. The computations were performed on resources provided by the Swedish National Infrastructure for Computing (SNIC) at NSC and HPC2N.

## Conflict of Interest

The authors declare no conflict of interest.

## Data Availability Statement

The data that support the findings of this study are available from the corresponding author upon reasonable request.

## Keywords

2;3-dihydrothieno[3, 4]dioxin-5-yl)thiophene, 4-b][1, 5-bis(2, electropolymerization, ETE-S, evolvable transistors, organic electrochemical transistors, silanes, synaptic transistors

Received: February 26, 2022
Revised: April 8, 2022
Published online: May 20, 2022